# Special Relativity (Lorentz Transformation) Follows from the Definition of Inertial Frames

Somajit Dey

*Department of Physics, University of Calcutta, 92, A.P.C Road, Kolkata-700009, India*
Email: sdphys_rs@caluniv.ac.in and dey.somajit@gmail.com

Besides the defining space-time symmetries (homogeneity and isotropy) of inertial frames, the derivation of Lorentz transformation requires postulating the principle of relativity and the existence of a finite speed limit. In this article, we point out that the existence of a finite speed limit can be readily inferred from the nature of allowed inertial frames. We also show that the principle of relativity can be obtained from the defining space-time symmetries of every inertial frame. Therefore, if the conventional definition of inertial frames is augmented properly, the special theory of relativity (Lorentz transformation) would follow from the definition of inertial frames.

## I. INTRODUCTION

Any frame with space-time homogeneity (STH) and spatial isotropy (SI) is called an inertial frame [1]. In Ref. [2] we suggested an extension to that definition by adding to the above symmetries the isotropy in time (time-reversal symmetry). Apart from these symmetries, the derivation of Lorentz transformation requires the postulate of principle of relativity (PR) and the existence of a finite speed limit in every inertial frame [3].

Existence of a finite speed limit is the only postulate among the above ones that distinguishes the Lorentz transformations from the Galilean ones [2, 4]. So the relativistic Lorentz transformation is qualitatively different from the non-relativistic Galilean one. However, it is commonly believed that relativistic effects become significant only for speeds comparable to that of light in vacuum $c$. Even if the world is relativistic, experimentally we cannot know it with certainty until we deal with such speeds. Hence, the concept of non-relativistic limit for speeds $v \ll c$. (In the non-relativistic limit, we ignore phenomena involving mass-energy conversion like nuclear decay or fission). But the existence of such a non-relativistic limit seems surprising when we consider the fundamental qualitative difference (viz. the existence and non-existence of a finite speed limit) between a relativistic and non-relativistic world. If the world was non-relativistic, it should be *qualitatively* different from what it is when it is relativistic, even in the $v \ll c$ limit. As we will argue in this article (Sec. II), *this difference lies in the non-existence of an inertial frame which is at rest relative to another inertial frame but not related to it by a Euclidean transformation (viz. translation and/or orthogonal linear transformation).* The existence of a finite speed limit in every inertial frame thus can readily be inferred (even for low or zero speeds) from the nature of the allowed inertial frames.

Another focus of this article is PR. PR states that all inertial frames are equivalent. Equivalence between two frames means that none of them is preferred to the other or, in other words, each is equally good for describing the laws of physics. Now, all the defining properties of a system, when taken together, are expected to characterise fully the system they define. It seems justified therefore, that the equivalence of inertial frames, i.e. PR, can be established from the universal defining properties of all inertial frames. In this article (Sec. III) we show that PR indeed follows from the defining symmetries of STH, SI in every inertial frame. After all, to say that no direction in space (SI) or point in space and time (STH) is preferred per se, indeed implies the necessary import of "equivalence". It stands to reason therefore, that the equivalence stated in PR might actually be a logical extension of these for the case of inter-frame relationships.

Our conclusions will be as follows. Since the existence of a finite speed limit is inferable from the nature of allowed inertial frames, it can be considered as a defining property of inertial frames. Also, since PR follows from the definition of inertial frames it is no more an independent postulate. The defining properties of inertial frames, viz. the homogeneity and isotropy of space and time and the existence of a finite speed limit, therefore, exhaust all the postulates of the special theory of relativity.

## II. EXISTENCE OF A FINITE SPEED LIMIT

Consider any two inertial frames with Cartesian coordinates, $S(x, y, z, t)$ and $S'(x', y', z', t')$. The three rectilinear space-coordinates and time are represented by their usual symbols. From STH, it follows that the most general transformation between $S$ and $S'$ is linear [5] viz.

$$\begin{pmatrix} x' \\ y' \\ z' \\ t' \end{pmatrix} = \mathbf{T} \begin{pmatrix} x \\ y \\ z \\ t \end{pmatrix} + \begin{pmatrix} o_x \\ o_y \\ o_z \\ o_t \end{pmatrix}, \qquad (1)$$

where $\mathbf{T}$ is a transformation matrix independent of $(x, y, z, t)$, and the right most column matrix is a constant dependent on the choice of origin. For points fixed in the $S'$ space, $\frac{dx'}{dt} = 0$, $\frac{dy'}{dt} = 0$, $\frac{dz'}{dt} = 0$ as seen from $S$. Hence, differentiating the first three rows of Eq. (1) with respect to $t$ we get

$$\begin{pmatrix} 0 \\ 0 \\ 0 \end{pmatrix} = \mathbf{\Gamma} \begin{pmatrix} \frac{dx}{dt} \\ \frac{dy}{dt} \\ \frac{dz}{dt} \end{pmatrix} + \begin{pmatrix} n_x \\ n_y \\ n_z \end{pmatrix}. \qquad (2)$$

where $\mathbf{\Gamma}$ is some $3 \times 3$ matrix and the rightmost column is constant. Assuming we can solve for $(dx/dt, \; dy/dt, \; dz/dt)^T$, Eq. (2) implies that $S'$ moves with a constant velocity with respect to $S$. Therefore, all inertial frames move with uniform velocity (null or non-null) relative to each other.

Now, let us look into the most general transformation between any two inertial frames $S$ and $S'$, that are at rest relative to each other. In this case, the rightmost column in Eq. (2) must be a null matrix. Therefore, choosing a common origin, the first 3 rows in Eq. (1) become

$$\begin{pmatrix} x' \\ y' \\ z' \end{pmatrix} = \mathbf{R} \begin{pmatrix} x \\ y \\ z \end{pmatrix}, \qquad (3)$$

where $\mathbf{R}$ is a transformation matrix independent of $(x, y, z, t)$. Since, $\Delta x = \Delta y = \Delta z = 0$ implies $\Delta x' = \Delta y' = \Delta z' = 0$, from the last row of Eq. (3) we get $\Delta t' = \alpha \Delta t$, $\alpha$ being some constant independent of $(x, y, z, t)$. This implies that by choosing proper units, the local times of the primed and unprimed frames can be set equal. Hence,

$$t' = t. \qquad (4)$$

Differentiating Eq. (3) with respect to Eq. (4), we find

$$\begin{pmatrix} dx'/dt' \\ dy'/dt' \\ dz'/dt' \end{pmatrix} = \mathbf{R} \begin{pmatrix} dx/dt \\ dy/dt \\ dz/dt \end{pmatrix}. \qquad (5)$$

Since **R** is independent of $(x, y, z, t)$, Eq. (5) implies that a free particle moving with uniform velocity in $S$ will also move with a uniform velocity in $S'$. Hence, $S'$ must be inertial and spatially isotropic if $S$ is so [1].

It is known that two inertial frames at rest relative to each other must be related by a rotation (if not a translation or reflection). **R** in Eq. (3), therefore, should be an orthogonal rotation matrix. This means that if we are in an inertial frame, looking at a circle, we cannot step into another frame at rest from which the circle looks like an ellipse, for example. However, we cannot prove from STH and SI of $S$ and $S'$, viz. their inertiality [1], that **R** must be an orthogonal matrix. This is because even if **R** is non-orthogonal, $S'$ and $S$ still satisfy STH and SI. The principle of relativity, another hallmark of inertial frames, also does not help in this regard. This is because it declares the similarity of natural laws in every inertial frame; but we proved that $S$ and $S'$ are both inertial, so forms of physical laws must be similar in them even if they are related by a non-orthogonal transformation. (If all the laws of physics involve only Cartesian tensors in order to be rotation invariant, they still remain invariant after the transformation $S \rightarrow S'$). Assuming the existence of rigid rods will not help too. (Note in passing that the existence of rigidity implies the non-existence of a finite speed limit). It may intuitively seem obvious that rotating a rigid metre stick in $S$ will look like a rotation of a rigid rod of *constant* length from $S'$. However, this may not be the case, since any non-orthogonality of **R** would imply that the rod will undergo length contraction/expansion anisotropically (i.e. changing length as it rotates) in $S'$. This however, does not violate the inherent SI of $S'$. The length of the rotating rod changes anisotropically in $S'$ only because the rod is seen to be rotating with uniform length from $S$ (and not from $S'$). Conversely, rotation of a rod of uniform length in $S'$ will look like rotation of a rod of variable length from $S$. There is no logical self-contradiction here. Firing identical balls with some mechanism (like springs) in every direction with identical speed is also not a frame-independent objective fact and is actually a variation of the above argument with rigid rods. Two identical balls thrown along the $X$ and $Y$ axes (say) in $S$ with the same speed may not seem to be moving with the same speed when seen from $S'$, for non-orthogonal spatial transformation between $S$ and $S'$. This is because, we can say that the mechanisms of the throws are the same for both the directions (hence the identical speeds) only when seen from $S$ and not from $S'$.

To elaborate, a non-orthogonal **R** means that the units of length in $S'$ do not look the same in all the directions when seen from $S$. If the unit of length in the $X$ direction (say) agrees with the unit of length of $S$, the unit of length in the $Y$ direction may not. Seen from $S$, the frame $S'$ thus looks anisotropic, but in itself , i.e. when seen from $S'$ itself, $S'$ is still isotropic. That this makes sense can be seen if we compare the analogous situation of two inertial frames moving along the $X$ axis of each other. Unit of length in one frame along the $X$ axis, when seen from the other frame, looks different from that along the $Y$ or $Z$ axis, but both frames are inherently isotropic.

If, however, there exists a finite speed limit in every inertial frame, **R** must be orthogonal. This can be argued as follows. Let the speed limit in $S$ be $V$ and that in $S'$ be $V'$ ($V'$ and $V$ need not necessarily be equal). It can be simply argued from Eq. (5) that a particle moving with the maximum speed in one frame must move with the maximum speed in the other frame too (see Appendix A). So motion with maximum speed is a frame-independent objective phenomenon. Now consider particles moving with maximum speed in all directions from the common origin of $S$ and $S'$. After any non-zero time interval $t = t'$, the particles will be situated on a sphere when seen from any of the frames. Hence, a sphere must transform into a sphere through the transformation $S \rightarrow S'$ viz. Eq. (3). This implies that **R** must be orthogonal. In other words, the existence of a finite speed limit in every inertial frame implies that two inertial frames at rest relative to each other must be related by a Euclidean transformation (translation and/or orthogonal linear transformation) in their space coordinates. This property of inertial frames, viz. their being related by Euclidean transformations alone when at rest, is sometimes taken as a postulate of special theory of relativity in order to derive Lorentz transformations [6]. As we saw, this postulate should be equivalent to the postulate of a finite speed limit in every frame.

All the above can be justified in still another way. We know that the relativistic Lorentz transformation transforms into non-relativistic Galilean transformation in the $c \rightarrow \infty$ (infinity) limit. In relativistic Minkowski space-time, the distance between two space-time points (events)

$$d = (\Delta \mathbf{r})^2 - c^2 (\Delta t)^2 \qquad (6)$$

remains invariant in all inertial frame transformations (above, $\Delta \mathbf{r}$ denotes the difference in position vectors of the two events). We can also say that $d / c^2$ is the metric that remains invariant under coordinate transformation. Note that for two inertial frames at rest to each other, this implies that the Euclidean metric (i.e. the spatial distance $\sqrt{(\Delta \mathbf{r})^2}$ ) must remain invariant in inter-frame transformation, since $\Delta t$ is the same for both the frames. Hence, these frames cannot be related by a non-Euclidean spatial transformation. In the non-relativistic case however,

$$\lim_{c \rightarrow \infty} (d / c^2) = -(\Delta t)^2 . \qquad (7)$$

This implies that in all inertial frame transformations only time *must* remain invariant. Note that the invariance of spatial distances (viz. the Euclidean metric $\sqrt{(\Delta \mathbf{r})^2}$ ) is no more implied as a logical necessity even for two inertial frames relatively at rest. Hence, the non-Euclidean spatial transformations remain valid possibilities for these non-relativistic frames. This is to say that even though our familiar Galilean transformations are such that two frames at rest must be related by a Euclidean transformation, the space-time symmetries (homogeneity and isotropy) of inertial frames alone fail to explain why the non-Euclidean transformations are ruled out.

We, therefore, find that the existence of a finite speed limit, as opposed to its non-existence, excludes the possibility of non-trivially different (i.e. not related by a rotation and/or reflection and/or translation) inertial frames at rest relative to each other. That there are no such frames is an immediate consequence of the world being relativistic, viz. the existence of a finite speed limit. Should the world be non-relativistic (i.e. without any finite speed limit), it would have been *qualitatively* different from what it is even in the $v << c$ limit. In other words, the only assumption distinguishing the Lorentz transformation from its Galilean counterpart, viz. the existence of a finite speed limit, manifests itself even in the $v << c$ limit by determining the structure of allowed inertial frames. Note, however, that the above arguments apply only for more than 1 dimensional space.

To appreciate the postulate of finite speed limit further, we refer the interested reader to the contribution by Drory [4].

### III. DERIVATION OF PRINCIPLE OF RELATIVITY (PR)

In what follows, we try to establish that any two frames, each of which satisfies STH and SI and has a finite speed limit, are equivalent i.e. they satisfy PR. However, we need to prove the following lemmas first. (Lemmas are not postulates; they can be proved.)

**Lemma 1: *The most general transformation between two inertial frames consists of a pure boost and a Euclidean transformation.***

**Proof:** We saw in Sec. II that two inertial frames move with a null or non-null velocity relative to each other. By a pure boost we mean the transformation that exists by virtue of a uniform relative velocity alone. Sec. II also tells that considering the existence of finite speed limit in each of them, two inertial frames at rest must be related to each other by a Euclidean transformation, viz.

translation and/or rotation. Let us take now any arbitrary inertial frame $B$ moving with a uniform velocity relative to another inertial frame $C$. Let a third frame $D$, related to $C$ by a pure boost, also move with the same velocity as $B$ with respect to $C$. Therefore, $B$ and $D$ are at rest relative to each other and hence, should transform into each other by a Euclidean transformation only. The most general transformation that takes $C$ to $B$ is thus composed of a pure boost ($C \rightarrow D$) and a Euclidean transformation ($D \rightarrow B$). Since we chose $C$ and $B$ arbitrarily, we have proved lemma 1.

**Remarks:** By SI and STH, two inertial frames related by a Euclidean transformation (spatial rotation/translation) are equivalent. Hence, to prove PR, it is sufficient to show that any two inertial frames $S$ and $S'$, related by a pure boost must be equivalent.

**Lemma 2:** *Seen from any inertial frame, all inertial frames related to it by a pure boost and moving with the same speed must be equivalent.*

**Proof:** Suppose the statement is false. Then a frame boosted in one direction would be preferred (i.e. non-equivalent) to a frame boosted similarly (i.e. with the same speed) in another direction. This would violate the SI of the original (unboosted) inertial frame. Hence, the statement in Lemma 2 must be true.

**Lemma 3:** *Two inertial frames moving with the same speed relative to a third must be equivalent.*

Proof: This follows from Lemma 1 and Lemma 2 when considered together.

Now we go on to prove the equivalence between two arbitrary inertial frames $S$ and $S'$ related by a pure boost. Let the speed of $S'$ relative to $S$ be $v$ along the positive $X$ axis. It can be argued that [2]

$$x' = -cx + cvt \qquad (8)$$

$$y' = ay - dvz \qquad (9)$$

$$z' = az + dvy \qquad (10)$$

$$t' = et + fvx \qquad (11)$$

where $c, a, d, e, f$ are all scalar functions of speed $v$. Using the time isotropy it can be shown that $d = 0$ [2]. But even if $d \neq 0$, we can make it 0 by rotating the $S'$ frame about the $X$ axis. By SI in $S'$, the new frame will be equivalent to $S'$. Hence, in the following we ignore $d$. From Eq. (8)-(11) it follows that,

$$\frac{dx'}{dt'} = \frac{-c\left(\dfrac{dx}{dt} - v\right)}{e + fv\dfrac{dx}{dt}}, \qquad (12)$$

$$\frac{dy'}{dt'} = \frac{a\dfrac{dy}{dt}}{e + fv\dfrac{dx}{dt}} \text{ and} \qquad (13)$$

$$\frac{dz'}{dt'} = \frac{a\dfrac{dz}{dt}}{e + fv\dfrac{dx}{dt}}. \qquad (14)$$

The speed of $S$ relative to $S'$ is $v' = \left|c\dfrac{v}{e}\right|$ along the $X'$ axis.

Consider now, the set $U$ of all inertial frames obtained from $S$ by pure boosts of speed $v$ in every direction possible. $S'$, therefore, belongs to $U$. So, atleast one element in $U$ is at rest relative to $S'$. Evidently however, there are frames in $U$ that move with speeds greater than $v'$ relative to $S'$. For example, consider the frames boosted along the $Y$ axis in $S$; when seen from $S'$, they move with speed greater than $v'$. In $U$, as seen from $S'$, therefore, we can always find frames with speeds continuously distributed in the range $[0, w]$ where $w > v'$. The continuity derives from the assumption that there exists a finite speed limit [Appendix B]. Hence, there must be some frame $S''$ in $U$, that moves with speed $v'$ relative to $S'$. $S''$, therefore, moves with the same speed as $S$ when seen from $S'$, and with the same speed as $S'$ when seen from $S$. By Lemma 3 therefore, $S''$ is equivalent both to $S'$ and $S$. By the transitive property of equivalence, $S$ and $S'$ thus must be equivalent. Hence, PR stands proved (see Remarks following Lemma 1).

The above strategy to prove PR works well for 2D and 3D space but not for 1D. In what follows, we prove PR from the same defining properties of inertial frames for 1D space. Note that Lemma 3 still remains valid for 1D space.

Consider an inertial frame $S$. Now consider another inertial frame $S'$, moving relative to $S$ with velocity $v$. Eq. (8) and (11), and hence, Eq. (12) also apply in this case. Relative velocity of $S$ with respect to $S'$ is $c\dfrac{v}{e}$. Consider now, another frame $S''$ moving relative to $S'$ with velocity equal and opposite to that of $S$, i.e. with velocity $-c\dfrac{v}{e}$. By Lemma 3, $S$ and $S''$ must be equivalent. Eq. (12) implies,

$$\frac{dx}{dt} = \frac{e\left(\dfrac{dx'}{dt'} - \dfrac{cv}{e}\right)}{-c - fv\dfrac{dx'}{dt'}}. \qquad (15)$$

$S''$, therefore, moves with velocity

$$v'' = \frac{2v}{1 - \dfrac{fv^2}{e}} \qquad (16)$$

relative to $S$. In Ref. [2] it is shown that $\left|\dfrac{e}{fv}\right|$ is greater than the finite speed limit in $S$ [7]. Hence, $\left|\dfrac{e}{fv}\right| > v \Rightarrow \left|\dfrac{fv^2}{e}\right| < 1$. (Note that this remains true even if $f = 0$, which implies there is no finite speed limit [2]). From Eq. (16) it is, therefore, seen that

$$\frac{dv''}{dv} = \frac{2\left(1 + \dfrac{fv^2}{e}\right)}{\left(1 - \dfrac{fv^2}{e}\right)^2} > 0 \qquad (17)$$

$v''$ is, therefore, a monotonically increasing continuous function of $v$ as long as $v$ is within the finite speed limit. This implies that $v''$ is a one-one function of $v$. (Also note that, $v'' = 0$ if and only if $v = 0$). For every allowed $v''$, we, therefore, can find a unique $v$. Hence for all possible inertial frame $S''$ (i.e. for all possible uniform velocity $v''$ relative to $S$), there will be an $S'$ with respect to which $S''$ and $S$ move with equal but opposite velocity and hence are equivalent by Lemma 3. Hence, PR in 1D is proved.

## IV. DEFINITION OF INERTIAL FRAMES

In light of the above discussion we suggest the following definition of inertial frames.

*Inertial frame is a reference frame having Euclidean space and time such that,*

*1) Space is homogeneous*[1]

*2) Time is homogeneous*[1]

*3) Space is isotropic*[1]

*4) Time is isotropic (i.e. Time reversal symmetry)*[2]

*5) There exists a finite speed limit* [Sec. II]

## V. JUSTIFICATIONS OF THE ABOVE DEFINITION

Reference frames prescribe coordinates (space and time) to describe natural phenomena. The above definition of an inertial frame (Sec. IV) makes the local space-time prescription possess the highest possible symmetry. (The conventional definition without time isotropy [1] is asymmetric in space and time). Even for 1D space, time isotropy is physically different from spatial isotropy (i.e. inversion symmetry for 1D space), since spatial inversion reverses velocities, accelerations and positions, while time reversal reverses only the velocities.

As we saw in Sec. III, the principle of relativity also follows from the above definition of inertial frames (Sec. IV).

This universal definition of inertial frames (Sec. IV), therefore, possesses or gives all the postulates required for a conventional derivation of Lorentz transformations [3]. In other words, inertial frames, when defined in such a way, must be related to each other by the elements of Poincare and Lorentz group. Having obtained this kinematical (Lorentz) transformation, we can then obtain other important aspects (e.g. $E = mc^2$ [8]) of the special theory of relativity.

## VI. THE SCALING PROBLEM

In order to keep things simple, we have not considered the scaling problem in all the above. We discuss it now. In Eq. (3), we saw that two inertial frames at rest relative to each other must be related by a matrix ( $\mathbf{R}$ ) transformation in space and a scaling ( $a$ ) in time. Although we kept $\alpha = 1$ (by changing the units of time) for simplicity, such an elimination of the scaling in time is not necessary for showing that a sphere in unprimed frame transforms into a sphere in the primed (Sec. II). Also, this 'sphere transforms to sphere' argument does not rule out the possibility of $\mathbf{R}$ being a scaled rotation matrix, viz. $\mathbf{R} = a\mathbf{O}$ where $a$ is the scaling constant and $\mathbf{O}$ is an orthogonal matrix. For purposes of simplicity, thus far, we have ignored the possibility of $a \neq 1$. Note that $a$ cannot always be made 1 by choosing proper length units. That $a \neq 1$ is non-trivial, can be appreciated as the following. Suppose we have two identical rigid rods of same length in an inertial frame $A$. Now we accelerate them arbitrarily yet differently from one another, such that ultimately they both move with the same velocity relative to $A$. Now, the rods are at rest relative to each other, yet $a \neq 1$ means that they may differ in their lengths. What more, this implies that the transformation between two inertial frames related by a boost (frame transformation by virtue of relative velocity only) may depend (in terms of scaling) on the actual physical process by which that boost is achieved, e.g. whether it is achieved by a uniform acceleration or a non-uniform one. To be more specific, we cannot say that $c, a, d, e, f$ in Eq. (8)-(11) must depend only on the final relative velocity of $S'$ with respect to $S$ ; they may as well depend on the specific history of the boost (e.g. the velocity-time graph made by the boost).

However, it can be shown that space and time scale identically, i.e. $\alpha = a$. First, note that the possibility of these scalings does not jeopardize the derivation of principle of relativity (Sec. III). The forms of natural laws are immune to such scalings since we can describe physics equally well (i.e. in the same form) in any units we choose. So principle of relativity or equivalence of inertial frames still holds in the face of these scalings. Now, an interval of space (i.e. length ( $l$ )) can be universally agreed upon by defining it as the length of a given rigid rod when brought to rest relative to the inertial observer. Specifying an interval of time (e.g. universal unit time ( $\tau$ )), however, unavoidably needs reference to the maximum speed limit ( $V$ ) intrinsic to the inertial observer ( $\tau = l/V$ ). Motivated by the equivalence of all inertial frames (principle of relativity), an inertial observer's most natural choice is to consider $V$ to be same in all inertial frames. So, if units of length differ in two frames, units of time must differ similarly, keeping $V$ the same in both of them. Hence, $\alpha = a$ and the frame-invariance of the finite speed limit.

Now we go on to show that $\alpha = a = 1$ (as we know from hindsight). This requires the concept of an infinitesimal boost. The boost that gives an inertial frame a relative velocity $d\mathbf{v}$ in an infinitesimal time $dt \to 0$ is called an infinitesimal boost. Since $d\mathbf{v} = \mathbf{a}dt$ , this boost can only be achieved by means of a uniform acceleration, $\mathbf{a}$ . Consider the frame $S$ infinitesimally boosted into $S'$ by a uniform acceleration $\mathbf{a}$ . From SI and STH, Eq. (8)-(11) follow [2] with $v$ replaced by $dv = |d\mathbf{v}|$ . Although, $c, a, d, e, f$ should be functions of $|\mathbf{a}|$ and $dt$ now, dimensional considerations show that they must occur as $|\mathbf{a}|dt = dv$ . To illustrate, consider the function $c$ . From Eq. (8) it must be dimensionless. Now the only quantities that go in the problem are $\mathbf{a}$ and $dt$ (by statement of the problem) and the finite speed limit $V$ (by defining property of inertial frames). $c$ will be non-trivially dimensionless only if it has arguments of the form $|\mathbf{a}|dt/V$ . Similarly, the transformation $S' \to S$ , depends on $d\mathbf{v}'$ viz. the relative velocity of $S$ as seen from the inertial primed frame. By reciprocity principle [2] (which remains valid by virtue of $\alpha = a$ ) $d\mathbf{v}' = -d\mathbf{v}$ . Coupled with the principle of relativity and the frame invariance of finite speed limit, this implies that the transformation $S \leftrightarrow S'$ is a Lorentz transformation [2]. Now, any finite boost can be considered as an infinite succession of infinitesimal boosts. Consider two arbitrary finite boost operators $F^1$ and $F^2$ , each giving a boost to final relative velocity $\mathbf{v}$ . Generally they should be related by a scaling of $a = \alpha$ viz. $F^1 = aF^2$ . However, as a succession of Lorentz transformations, any finite boost must conserve the Minkowski metric $(\Delta x)^2 - c^2(\Delta t)^2$ . Since $F^1$ and $F^2$ both conserve this metric individually, it follows that $a = 1 = \alpha$ .

The resolution of the scaling problem, therefore, required the assumption that every finite boost can be considered as an infinite succession of infinitesimal boosts. The reader is urged to compare this with the statement in Goldstein [9]:

"Consider a particle moving in the laboratory system with a velocity $\mathbf{v}$ that is not constant...We imagine an infinite number of inertial systems moving uniformly relative to the laboratory system, one of which instantaneously matches the velocity of the particle...The particle is thus instantaneously at rest in an inertial system that can be connected to the laboratory system by a Lorentz transformation. It is assumed that this Lorentz transformation will also describe the properties of the particle and its true rest system as seen from the laboratory system."

## VII. CONCLUSION

The foundations of special theory of relativity have undergone much scrutiny, starting as early as 1910 [10]. The minimal axioms required to rigorously derive Lorentz transformations have been sought and debated vigorously engendering a wide variety of derivations from a corresponding set of postulates. Drory [4] gives a nice account in his introduction. A glimpse of the sheer volume of the related literature can be found in the rich bibliography of Ref. [11]. In view of this, the sole goal of this work and its previous instalment [2] has been to point out some of the corners and unifying concepts that remained overlooked so far. All in all, we hope to have achieved something of significance towards tidying up the foundations of special relativity. In order to feed new perspectives effectively to a vision adapted to (and sometimes blinded by) conventional wisdom, we had to proceed step by step (for example the scaling problem in Sec. VI was not discussed until the very end). To clarify the true content of the present work a summary is in order.

When inertial frames are defined as in Sec. IV, the principle of relativity follows as a logical conclusion. Also, the defined existence of a finite speed limit in every inertial frame rules out the existence of two inertial frames at rest related by a non-Euclidean spatial transformation. The finite speed limit in every frame is taken to have the same value (since no inertial frame is preferred due to the principle of relativity). For a given universal choice of length unit, the unit of time then follows identically in every frame from that universal speed limit and the chosen unit of length. From space-time homogeneity, spatial isotropy, time isotropy, principle of relativity, existence and frame-invariance of a finite speed limit, Lorentz transformations for infinitesimal boosts are uniquely obtained (Ref. [2] and Sec. VI). Since all the above properties either exist in or follow from the definition of inertial frames as given here, Lorentz transformation for an infinitesimal boost actually becomes a logical conclusion of the inertial frames. Since, any finite boost is a succession of infinitesimal boosts, space-time intervals in boosted frames cannot depend on the specifics of the process of boosting.

If we conceive inertial frames as in Sec. IV and see finite boosts as an infinite succession of infinitesimal boosts, then inertial frames must be related by transformations of the Lorentz and Poincare group. Given the transformation rules, special relativity as a theory will then follow.

## APPENDIX A

Eq. (5) can be written as

$$v'\hat{\mathbf{u}}' = v\mathbf{R}\hat{\mathbf{u}} \,, \tag{18}$$

where $\hat{\mathbf{u}}$ denotes the unit column vector along a velocity in the unprimed frame and $v$ is the scalar magnitude of that velocity. Similarly for the primed frame. Transposing Eq. (18) and multiplying with itself,

$$(v')^2 \hat{\mathbf{u}}'^T \hat{\mathbf{u}}' = (v)^2 \hat{\mathbf{u}}^T \mathbf{R}^T \mathbf{R}\hat{\mathbf{u}} \,. \tag{19}$$

For constant $\hat{\mathbf{u}}$ and $\hat{\mathbf{u}}'$, $v'$ therefore increases with $v$ and vice versa. The maximum speed limit of $S$ is $V$. Suppose from Eq. (19) that this gives a speed $W'$ in the primed frame with $W' < V'$. From Eq. (19) again, let the maximum speed limit in $S'$, viz. $V'$ give the speed $W$ in the unprimed frame. Then $W > V$ which is impossible. Hence, our assumption $W' < V'$ must be invalid. This implies that a speed limit in one frame must map to that in the other.

## APPENDIX B

Consider, without loss of generality, only the frames boosted in the $X - Y$ plane. Then a frame boosted at an angle $\theta$ with the positive $X$ axis has the velocity components: $\frac{dx}{dt} = v\cos\theta$ and $\frac{dy}{dt} = v\sin\theta$. By Eq. (12)-(13), the speed of this boosted frame as seen from $S'$ is

$$v'_\theta = \sqrt{\left(\frac{-c(v\cos\theta - v)}{e + fv^2\cos\theta}\right)^2 + \left(\frac{av\sin\theta}{e + fv^2\cos\theta}\right)^2} \,. \tag{20}$$

$v'_\theta$ is a differentiable and hence continuous function of $\theta$ for $\left|\frac{fv^2}{e}\right| < 1$. In Ref. [2] it is shown that $\left|\frac{e}{fv}\right|$ is greater than the finite speed limit in $S$ [7]. Hence, $\left|\frac{e}{fv}\right| > v \Rightarrow \left|\frac{fv^2}{e}\right| < 1$. (Note that this remains true even for $f = 0$, viz. when there is no finite speed limit [2]).